# AI-based Automatic Segmentation of Prostate on Multi-modality Images: A Review

Rui Jin[a, 1], Derun Li[b, 1], Dehui Xiang[c, 1], Lei Zhang[d], Hailing Zhou[e], Fei Shi[c], Weifang Zhu[c], Jing Cai[f], Tao Peng[a, f, g *], Xinjian Chen[c, h, *]

[a] School of Future Science and Engineering, Soochow University, Suzhou, China
[b] Department of Urology, Peking University First Hospital, Xicheng District, Beijing, 100034, China
[c] MIPAV Lab, the School of Electronic and Information Engineering, Soochow University, Jiangsu, China
[d] Graduate Program of Medical Physics and Data Science Research Center, Duke Kunshan University, Kunshan, Jiangsu, 215316, China
[e] Department of Mechanical Engineering and Product Design Engineering, Swinburne University of Technology, Melbourne, Australia
[f] Department of Health Technology and Informatics, The Hong Kong Polytechnic University, Hong Kong, China
[g] Department of Radiation Oncology, UT Southwestern Medical Center, Dallas, Texas, USA
[h] The State Key Laboratory of Radiation Medicine and Protection, Soochow University, Jiangsu, China

**Acknowledgment:**

The authors acknowledge the project funded by China Postdoctoral Science Foundation (Certificate Number: 2023M742568).

[1] These authors contributed equally to this study.

[*] Corresponding author:
*E-mail* addresses: xjchen@suda.edu.cn (X.J. Chen), sdpengtao401@gmail.com (T. Peng)

# AI-based Automatic Segmentation of Prostate on Multi-modality Images: A Review

*Abstract*—**Prostate cancer represents a major threat to health. Early detection is vital in reducing the mortality rate among prostate cancer patients. One approach involves using multi-modality (CT, MRI, US, etc.) computer-aided diagnosis (CAD) systems for the prostate region. However, prostate segmentation is challenging due to imperfections in the images and the prostate's complex tissue structure. The advent of precision medicine and a significant increase in clinical capacity have spurred the need for various data-driven tasks in the field of medical imaging. Recently, numerous machine learning and data mining tools have been integrated into various medical areas, including image segmentation. This article proposes a new classification method that differentiates supervision types, either in number or kind, during the training phase. Subsequently, we conducted a survey on artificial intelligence (AI)-based automatic prostate segmentation methods, examining the advantages and limitations of each. Additionally, we introduce variants of evaluation metrics for the verification and performance assessment of the segmentation method and summarize the current challenges. Finally, future research directions and development trends are discussed, reflecting the outcomes of our literature survey, suggesting high-precision detection and treatment of prostate cancer as a promising avenue.**

*Impact Statement* — **Prostate cancer remains a significant health threat, necessitating early detection to reduce mortality rates. Current diagnostic approaches, particularly in medical imaging, face challenges in accurate prostate segmentation due to the organ's complex structure and imaging imperfections. Our review paper delves into AI-based automatic segmentation methods across various imaging modalities (CT, MRI, US), emphasizing their strengths, weaknesses, and the specific challenges they address. By classifying segmentation techniques based on supervision levels, we provide a comprehensive analysis of their performance and evaluation metrics. Our insights pave the way for future research, highlighting the potential of high-precision detection and treatment for prostate cancer. This advancement promises to enhance diagnostic accuracy, reduce dependence on physician expertise, and improve patient outcomes. As AI continues to evolve, its integration into prostate segmentation heralds a new era in medical imaging, offering more efficient and reliable diagnostic tools.**

## I. Introduction

PROSTATE cancer is the second most prevalent malignancy and the fifth leading cause of cancer-related mortality in men as of 2020. Remarkably, it is the most frequently diagnosed cancer in over half of the world's nations (112 out of 185 countries) [1]. Studies highlight the effectiveness of screening protocols in significantly reducing prostate cancer mortality rates [2]. Therefore, early detection of prostate cancer is crucial. With the rapid advancement of medical imaging technology, various imaging modalities, including computed tomography (CT), magnetic resonance imaging (MRI), and ultrasound (US), are critical in facilitating early disease diagnosis. The field of medical image processing has emerged as a key research area, emphasizing the importance of medical imaging. The increasing volume of image data, analysis of functional imaging data, and widespread use of complex and time-intensive techniques have markedly improved the precision of disease diagnosis. However, this advancement has also increased the complexity of disease diagnosis and the dependence on physician expertise [3]. Thus, it is essential to pursue relevant initiatives with computational methods.

The computer-aided design (CAD) system, utilizing computer technology and software tools, provides a comprehensive framework for aiding in the design, drafting, and analysis of products in fields such as engineering and manufacturing. In medical imaging, CAD is divided into computer-aided detection (CADe) and computer-aided diagnosis (CADx), differentiated by their specific goals [4]. A typical CAD system includes image preprocessing, region of interest definition, feature extraction, and subsequent selection and classification processes [5]. Integrating CAD technology into radiologists' workflows significantly improves interpretation speed and diagnostic accuracy, especially in diagnosing diseases like prostate cancer. Intelligent diagnostic outcomes offer valuable insights into the diagnostic process, enhancing both objectivity and efficiency. Yet, CAD systems face challenges due to the complexity and diversity of medical images, limited availability of annotated training data, and models' lack of interpretability. Thus, more work is needed to improve the accuracy of CAD systems.

Prostate segmentation is particularly valuable in prostate cancer diagnosis, treatment, and surgical procedures [6–8]. During prostate cancer diagnosis and treatment, segmenting the prostate helps physicians more accurately determine the cancer's location, size, and distribution, enabling a more effective treatment plan formulation. Additionally, prostate segmentation is key in monitoring the progression of prostate cancer, including factors like tumor size, location, and invasion depth [9]. This assessment aids in creating a more strategic treatment approach. Moreover, precise prostate segmentation enables accurate determination of the prostate's location and size, improving surgical precision and reducing associated risks.

Prostate image segmentation utilizes three methods: manual, semi-automatic, and automatic. Manual segmentation, though the simplest and most accurate, suffers from being time-consuming and has drawbacks such as significant subjectivity, limited repeatability, and difficulty in achieving three-dimensional segmentation [10]. With manual segmentation falling short of contemporary medical needs, there is a pressing need for automatic segmentation to better support medical diagnosis. The rise of precision medicine, increased clinical demand, and growing investment in AI have led to the widespread adoption of AI technology, particularly machine learning and data mining, in various medical areas, notably image segmentation.

This paper provides a detailed summary and classification of prostate segmentation techniques across three modes. Its goal is to clarify the correlations and differences among these techniques, thereby guiding future research and development in the field. Additionally, it suggests directions for subsequent research efforts. The quantitative evaluation of prostate segmentation methods is hampered by a lack of publicly available datasets, open-source software, and standardized evaluation indicators [11]. Thus, this paper proposes a framework for the verification and performance evaluation of these methods. Fig. 1 outlines the article's structure.

In summary, the primary contributions of this paper include:

- A detailed comparative analysis of multimodal imaging's strengths and weaknesses for prostate segmentation, addressing specific challenges of CT, MRI, and US.
- Presenting an innovative classification of AI-based models/algorithms for prostate image segmentation across various modes.
- Exploring challenges in the verification and performance evaluation of segmentation methods, including relevant evaluation indicators.
- Offering insights into future AI-based prostate image segmentation research

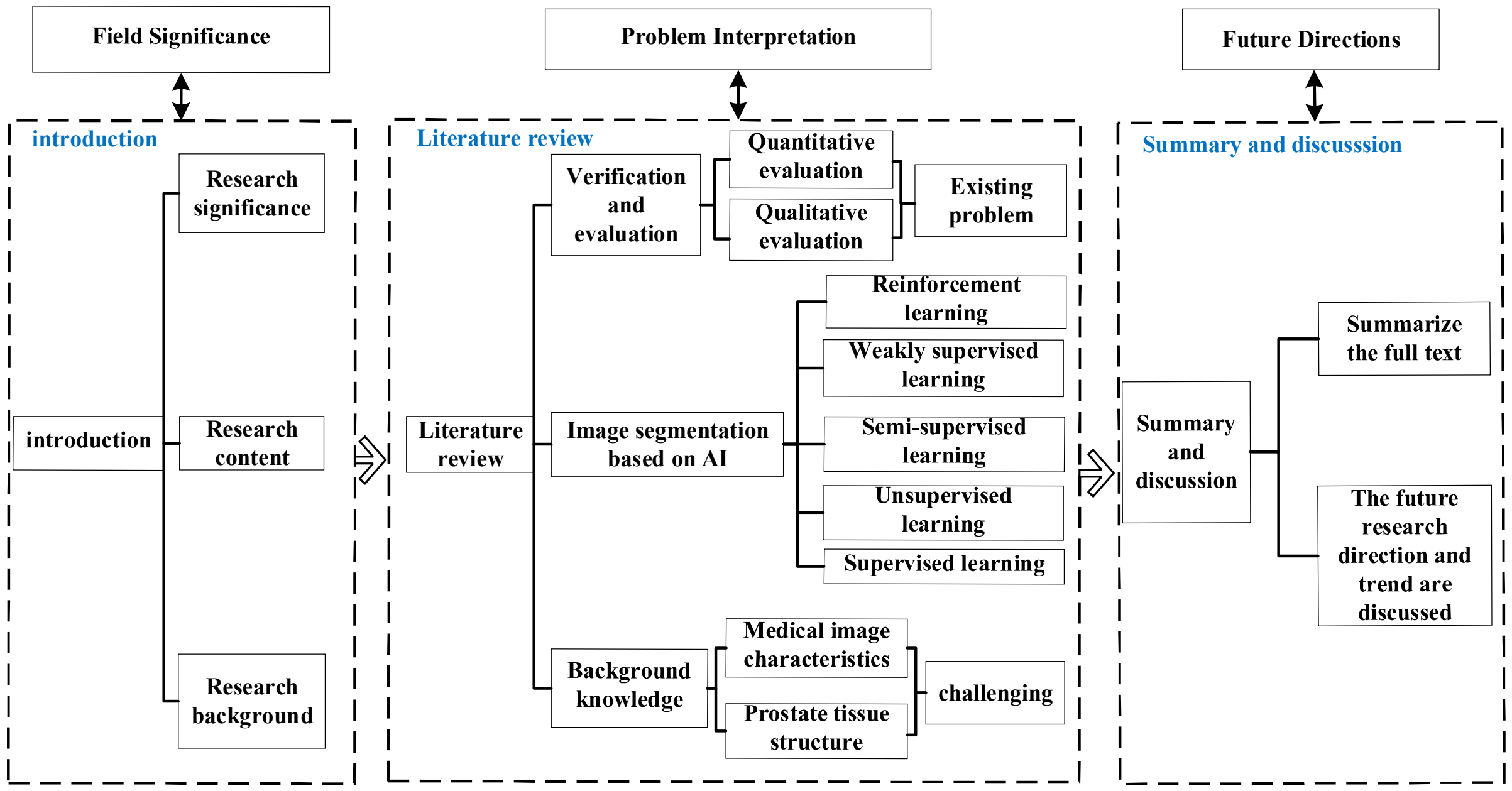

Fig. 1. Context diagram of article structure.

## II. Characteristics and Significance of Medical Image Segmentation

Standard diagnostic methods for prostate assessment include CT, MRI, and US [12]. This paper predominantly focuses on the unique characteristics of TRUS, MRI, and CT for their clinical use in automatic prostate segmentation. Fig. 2 [13], [14] succinctly summarizes the limitations of conventional prostate image segmentation methods.

### A. *Structure of prostate tissue*

Located at the bladder's neck and surrounding the urethra, the prostate is the largest accessory gland in men. It is encased in a thin layer of elastic fiber tissue, giving it an unlobulated appearance. Axially, the gland can be round, oval, or triangular. An inward extension of its capsule, made of fibrous elastic tissue, divides the prostate into several lobes: the anterior, posterior, middle, and two lateral lobes [15].Histologically, the prostate is divided into the anterior fibromuscular stroma, peripheral zone, central zone, transitional zone, and peri-urethral zone [16].

The prostate's intricate shape and structure, with various regions and types, increase the difficulty of segmentation [17]. Its inherent variability leads to morphological and structural differences across images, affecting the stability of segmentation algorithms. Moreover, the presence of artifacts, pseudo-structures, and image quality issues, including noise and other artifacts [18] , compromises segmentation accuracy.

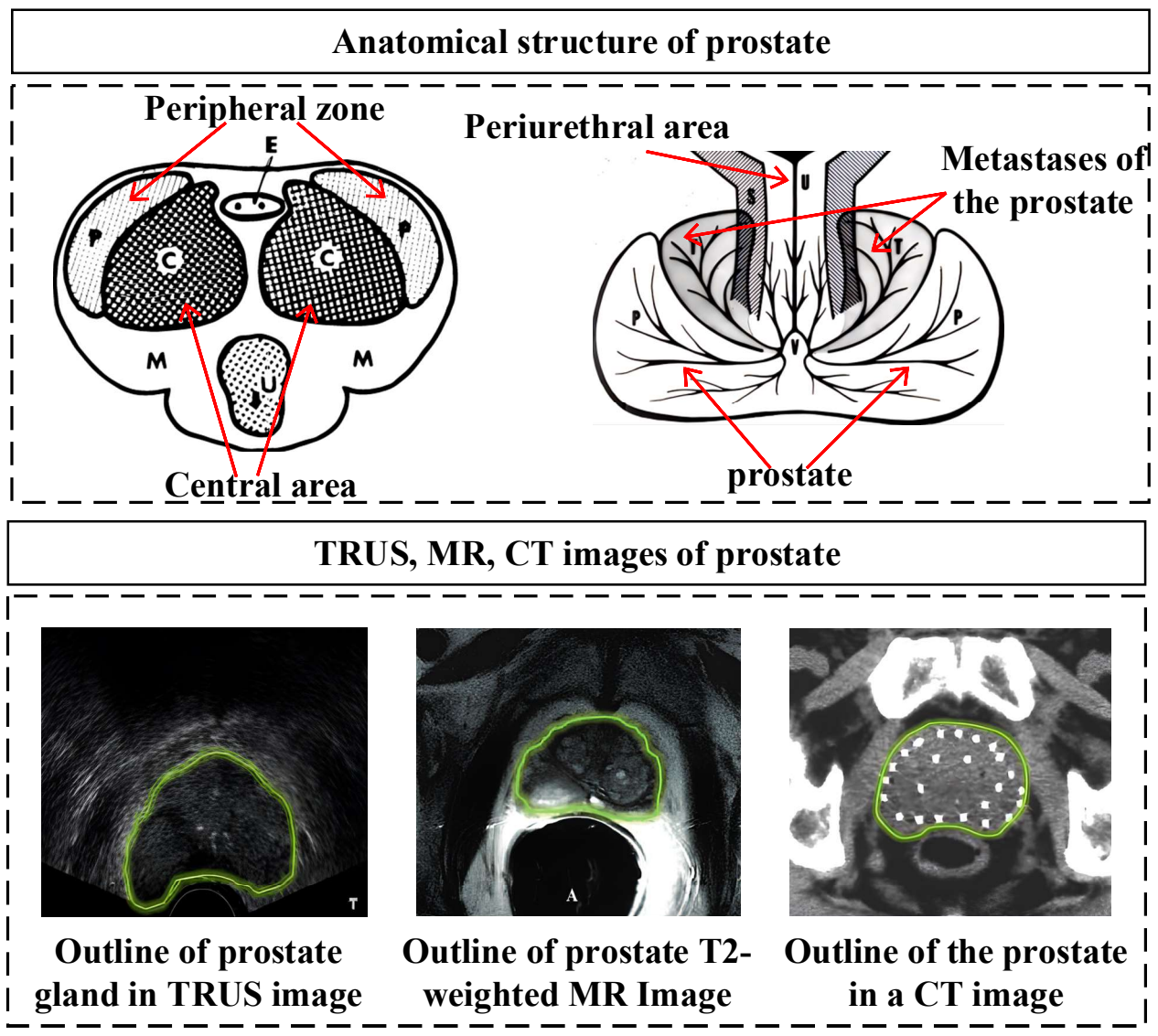

Fig. 2. Anatomical Structure and Contour of the Prostate[13], [14].

Uneven intensity distribution and potential pathological changes also challenge segmentation precision. Therefore, in processing prostate images, it is vital to use appropriate preprocessing and segmentation methods to overcome these challenges, ensuring accurate and reliable results [19].

*B. Medical imaging and prostate image segmentation*

This section focuses on the characteristics of three imaging techniques: US, MRI, and CT, as shown in Fig. 3.

| | **Advantage** | **Disadvantage** |
|---|---|---|
| **US** | • **No radiation, harmless to human body**<br>• **Low cost**<br>• **portability**<br>• **Real-time imaging** | • **Low contrast**<br>• **Low-resolution ancer**<br>• **There are spots, microcalcification, artifacts**<br>• **Difficult staging of cancer** |
| **MRI** | • **High definition of soft tissue structure**<br>• **Non-ionizing radiation**<br>• **It can scan any section**<br>• **Multi-channel images with varying contrast can be obtained** | • **expensive**<br>• **It is difficult to achieve real-time imaging**<br>• **Not portable**<br>• **There are certain requirements on the patient's body** |
| **CT** | • **Cross-section image thickness is accurate**<br>• **The cross-sectional image is clear**<br>• **Cross section image density resolution is high**<br>• **Fast imaging**<br>• **High quality 3D images can be generated**<br>• **It is suitable for different patients and has a wide range of applicability** | • **expensive**<br>• **It is difficult to achieve real-time imaging**<br>• **Not portable**<br>• **Soft tissue contrast difference**<br>• **There is radiation, which is harmful to human health**<br>• **Difficult staging of cancer** |

Fig. 3. Summary of advantages and disadvantages

*1) US prostate segmentation*

Medical US is a diagnostic imaging technology that employs sound waves to visualize soft tissues, such as muscles and internal organs [20]. US technology is widely used in medicine, favored for its non-invasive nature, cost-effectiveness, and real-time imaging capabilities [21]. Of these, TRUS is particularly used to obtain prostate images and in clinical settings, manual delineation of lesion areas in TRUS prostate images is often the standard for segmentation [22].TRUS is a key technique in cancer diagnosis and treatment, especially for visualizing pelvic organs [23]. Using a high-frequency transrectal probe close to these organs produces clear, high-resolution images. This method allows for the visualization of the uterus, endometrium, ovary, prostate, seminal vesicle gland, and rectal region, aiding in the detection of minor lesions.

However, due to the inherent limitations of its imaging principle, ultrasound images suffer from low resolution, high speckle noise density, low contrast, and artifacts [24], [25], resulting in uneven intensity within the prostate and blurred or missing boundaries with adjacent organs[26]. These issues hinder computer-assisted automatic or semi-automatic prostate segmentation. To overcome these challenges, various techniques are employed, such as adaptive selection of the principal curve and a smooth mathematical model [27], Auto-ProSeg [28], and semi-automatic segmentation from US images with machine learning and principal curves, supported by interpretable mathematical models [26]. Proposed solutions include semi-automatic or fully automatic segmentation algorithm models like H-SegMed [29], designed to accurately segment prostate US images.

*2) MRI prostate segmentation*

MRI is a powerful, non-invasive medical imaging technique. It uses strong magnetic fields, magnetic field gradients, and radio waves to create detailed internal organ images. MRI's primary advantages include high clarity of soft tissue structure, no ionizing radiation, the ability to scan any body section, and multi-channel image acquisition with variable contrast through different pulse sequences [30]. Therefore, MRI is invaluable for anatomical and functional studies of various body organs [31] and is a preferred method for tumor analysis, revealing location, size, shape, and intratumoral necrosis. Whole-body MRI provides detailed insights into the brain, liver, chest, abdomen, and pelvis, aiding doctors in diagnosis, examination, and treatment.

However, MRI has its limitations, such as higher costs, longer scanning times, more artifacts, and specific patient restrictions [32], which limit MRI's widespread use. While prostate MRI segmentation presents challenges, particularly in comparison to CT and US images, its superior soft tissue contrast yields high clarity in human soft tissue structures [33]. However, the prostate's complex structure, low contrast in adjacent regions of interest, blurred organ and tissue boundaries, and MRI artifacts make recognizing and extracting prostate features difficult, thus complicating accurate prostate segmentation. To address these issues, various solutions have been developed, such as the new human interaction-based semi-automatic prostate segmentation [34], the PC-SNet split network [35], and other semi-automatic or fully automatic segmentation algorithms/models.

*3) CT prostate segmentation*

CT [36] relies on the differential X-ray absorption and transmittance of human tissues, combined with the data from highly sensitive instruments and electronic computers, to generate cross-sectional or three-dimensional images of the body. This technique is effective in detecting small lesions in various body parts. CT imaging is widely used in the early diagnosis and screening of diseases in the brain, liver, chest, abdomen, pelvis, and spine, and for CT angiography. A CT image provides a cross-sectional view and is a staple in medical imaging. For a complete visualization of an organ, a series of consecutive cross-sectional images is necessary. These images offer precise layer thickness, high clarity, and density resolution, enhancing the visualization of organs comprising soft tissue[37].

However, prostate CT segmentation faces challenges. Firstly, the CT image's poor soft tissue resolution results in inadequate contrast between the prostate and surrounding tissues, impeding the use of boundary information [38]. Secondly, irregular and unpredictable prostate movement, along with variations caused by intestinal gas, leads to significant prostate position and shape changes on different treatment days. To overcome these issues, various models/algorithms have been proposed, including automatic segmentation for adaptive radiotherapy of the prostate [39] and semi-automatic segmentation based on coupled feature representation and spatial constraint direct push Lasso [40].

## III. AI-BASED IMAGE SEGMENTATION METHODS

With the advancement of precision medicine, increased clinical demand, and a boom in AI investments, AI technology, particularly machine learning and data mining, has significantly impacted various medical treatment fields. The concept of "AI+ medicine" holds broad prospects [41]. Currently, medical imaging is a prime application area for AI in the medical field. Integrating AI technology into medical image segmentation plays a key role in improving efficiency, reducing time, minimizing subjective deviation, and relieving doctors from the labor-intensive task of image segmentation [42]. As a subset of AI, machine learning, and particularly its branch, deep learning, has become a research focus in medical imaging due to its ability to process large amounts of data [43]. Recent developments in hardware have facilitated breakthroughs in deep learning methods, leading to a significant expansion in the research field based on deep learning and machine learning [44].

Contrasting with traditional rule-based algorithms, machine learning-based algorithms utilize a wealth of new data, continuously improving and learning over time without predefined programming [45]. Based on the amount of labeled data, machine learning is often categorized into supervised, unsupervised, semi-supervised, and weakly supervised learning. Supervised learning requires extensive pixel-level labeled data to train deep neural networks, such as fully convolutional network (FCN) [46], u-shaped network (U-Net) [47], mask region-based convolutional neural network (Mask R-CNN) [48]. Weakly supervised learning, on the other hand, only needs minimally labeled data like image-level tags, bounding boxes, or points. It generates pixel-level segmentation results through methods like attention mechanisms, multi-scale fusion, or adaptive thresholding, examples being multiple instance learning (MIL) [49] and weakly supervised semantic segmentation learning (WSSL) [50]. Semi-supervised learning requires a small amount of labeled data and a larger amount of unlabeled data, improving model generalization through self-training, adversarial learning, or data augmentation, as seen in Mean Teacher [51], MixMatch [52], and CutMix [53]. Unsupervised learning trains models using unlabeled data, clustering or generating based on image features or prior knowledge, such as K-means clustering (K-means) [54], gaussian mixture model (GMM) [55], and generative adversarial network (GAN) [56]. Reinforcement learning (RL) involves an agent interacting with an environment, learning to maximize rewards through "trial and error", exemplified by Deep Q-Network [57] and Actor-Critic [58].

Thus, AI-based image segmentation is classified into five types based on the level and nature of supervision during training: supervised learning, unsupervised learning, semi-supervised learning, weakly supervised learning, and reinforcement learning (RL). Fig. 4 depicts a classification scheme for AI-based image segmentation models

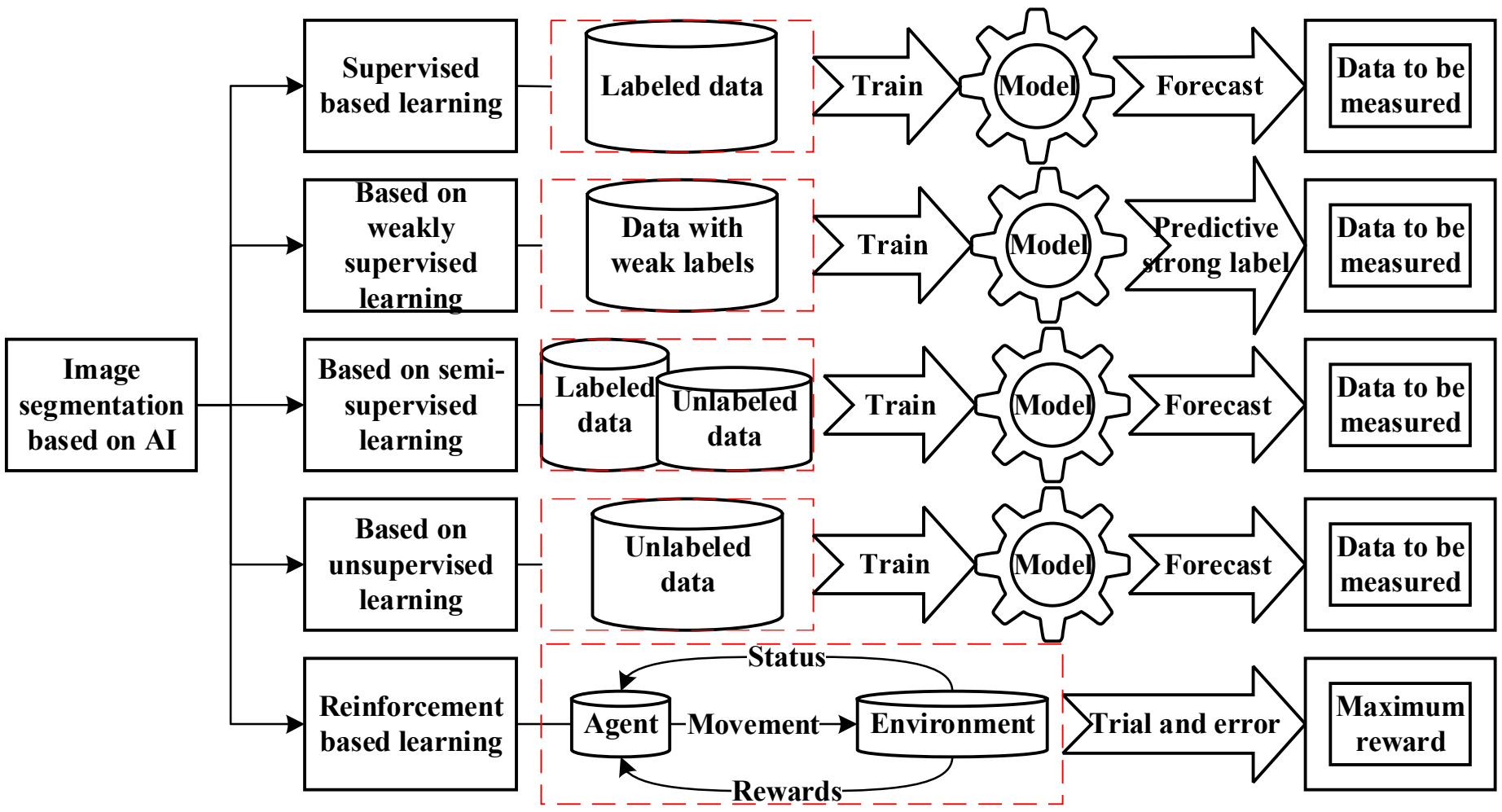

Fig. 4. Classification of AI-based image.

### A. *Supervised learning-based medical image segmentation*

#### 1) *Description of supervised learning*

Among various learning methods, supervised learning is most prevalent in radiology. It requires labeled training data, utilizes these labels to predict outcomes, and then compares predictions with actual results to identify errors and enhance the model [59]. Supervised segmentation methods incorporate prior knowledge through training samples, with techniques like support vector machines, random forests, and k-nearest neighbor clustering proving robust over the past [60]–[65]decade.

In medical image segmentation, where precision is crucial, supervised learning is the dominant approach. It offers several advantages [66]. Firstly, the learning process is straightforward and accurate, as it trains models using known labeled data. Supervised learning allows for result control by adjusting model parameters and optimizing algorithms, making it highly controllable. Additionally, it can automate medical image segmentation, minimizing human intervention and enhancing segmentation efficiency, thus enabling the automatic segmentation of medical images.

However, supervised learning's limitations are also evident. It relies on known data and corresponding labels, yet acquiring a substantial amount of labeled medical image data for clinical applications is challenging due to the scarcity of medical resources [66]. The organization of annotation data remains a significant bottleneck in the broader clinical application of supervised learning[63]. Challenges in prostate image segmentation using supervised learning include boundary incompleteness, blurring, and non-uniformity in US images [67], acquisition, representation, and updating of shape prior knowledge [68], and issues related to the registration, fusion, and enhancement of multimodal images [69]. To overcome these challenges, several innovative approaches have been proposed: 1) Utilizing semi-supervised, unsupervised, or weakly supervised learning to reduce dependence on labeled data; 2) Implementing attention mechanisms, multi-scale structures, or multi-viewpoint fusion to improve boundary information; 3) Constructing shape prior models through GAN [56], variational auto-encoder (VAE) [70], or dictionary learning [71]; 4) Processing multimodal images with multi-task learning [72], cross-domain transfer learning [73], or domain adaptation [74].

*2) Applications in supervised learning-based prostate segmentation*

In recent years, supervised learning methods, such as random forest and support vector machines, have gained widespread adoption in medical image segmentation. Inspired by combining supervised learning with a decision forest for medical image classification challenges, Ghose et al [75]. developed a probabilistic classification model using a decision forest for MRI prostate automatic segmentation. This method employs a random forest classification model before feature extraction to accommodate variations in the prostate's appearance. The posterior probability approach for identifying the prostate region is both simple and effective. The random forest classification framework is illustrated in Fig. 5 (a). Furthermore, Ghost et al. [76] introduced a supervised learning model using a random forest for the automatic initialization and propagation of statistical shape and appearance models, facilitating the automatic segmentation of real TRUS prostate. This method's accuracy and robustness have been validated experimentally, although the effectiveness on base and vertex slices was to be confirmed in the subsequent year. The segmentation accuracy of the model on base and vertex slices was later verified [67].

In scenarios with large datasets, graph-based image segmentation methods have proven effective, making them a popular choice in image segmentation [75]. Cheng et al. [77] developed a supervised learning framework combining the integrates the graph-based active appearance model (AAM) and support vector machine (SVM) model. This framework, illustrated in Fig. 5 (b), enables automatic segmentation of prostate MRI. The method was cross-validated on 40 MRI datasets, achieving an average segmentation accuracy of nearly 90%. Additionally, Cheng et al. [78] proposed a model that integrates a graph-based AAM with deep learning, achieving more precise automatic segmentation of prostate MRI using an adaptive graph-based AAM model and deep learning. Concurrently, Shi et al. [40] introduced a semi-automatic prostate segmentation method for CT images using coupled feature representation and spatially constrained direct push Lasso, leveraging prostate shape information from images to produce the final segmentation result through multi-atlas-based label fusion.

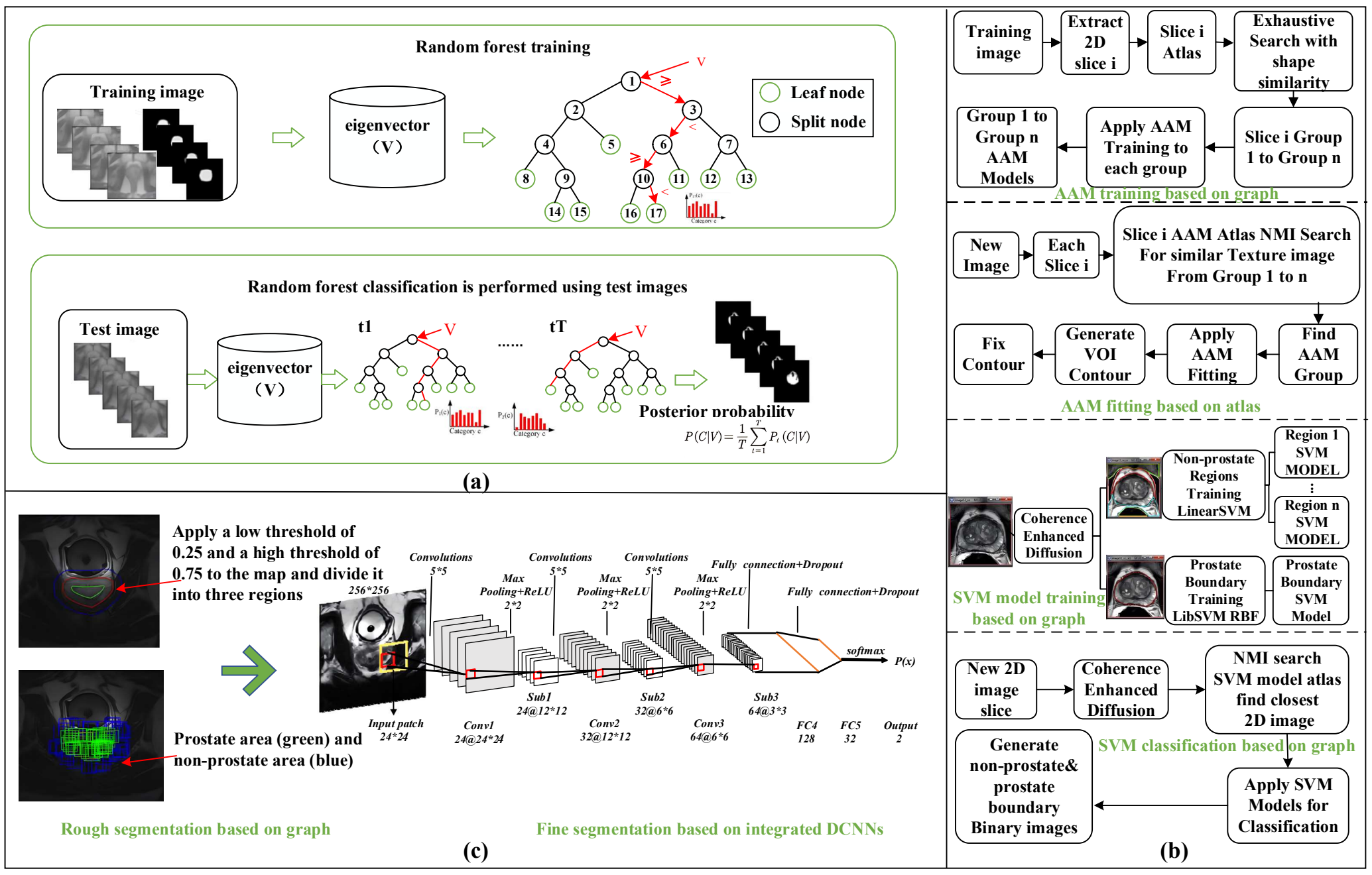

Fig. 5. Image segmentation method based on supervised learning. (a) Stochastic forest classification framework[75]; (b) Supervised learning framework based on AAM and SVM models [77]; (c) Coarse-to-fine segmentation strategy based on graphs and integrated deep Convolutional neural networks [79].

Owing to deep learning's excellence in computer vision, depth-supervised learning is increasingly employed in image segmentation. Jia et al. [79] proposed a coarse-to-fine segmentation strategy using integrated deep convolution neural networks (DCNNs) for prostate MRI segmentation. The method is divided into two stages: coarse segmentation based on graph and refinement using integrated DCNNs, as depicted in Fig. 5 (c). Additionally, Tian et al. [80] developed a deep full convolutional neural network (FCN) model (PSNet) for automatic prostate segmentation. Wang et al. introduced a method for automatic prostate segmentation on volume CT images [81] and MRI [82], [83] using a 3D depth-supervised expanding FCN based on depth FCN. However, manual segmentation remains a time-consuming and challenging intraoperative process in needle-based diagnosis and treatment of prostate cancer, given clinical diversity. To develop a universal algorithm for needle-based prostate cancer surgery, Orlando et al. [82] created a method using supervised deep learning with an enhanced U-Net for segmenting the prostate in 3DTRUS images from various devices. Moreover, Lei et al. [84] developed a multi-directional depth-supervised learning method for automatic prostate segmentation in US-guided radiotherapy, integrating a 3D monitoring mechanism into V-Net to address the challenge of training deep networks with limited data. Fig. 5 demonstrates an image segmentation method based on supervised learning.

Deep learning, despite its utility in prostate image segmentation, faces challenges including high computational and memory demands, as well as limited robustness and interpretability. A hybrid approach, combining traditional methods with deep learning techniques, leverages the strengths of each to offset their respective limitations, thus enhancing image segmentation performance. Peng et al. [85] developed an innovative hybrid method that merges an improved principal curve-based approach with an evolutionary neural network for segmenting prostate US images. This method achieved superior segmentation results and efficiency compared to deep learning alone and other hybrid methods, as depicted in Fig. 6 (a). Building on this, Peng et al. [27] introduced a combination of an adaptive selection principal curve model, an enhanced neural network, and an interpretable mathematical map function of the smooth boundary for US prostate image segmentation. This approach was thoroughly evaluated across multiple datasets and various shapes, showing improved segmentation results as illustrated in Fig. 6 (b). Fig. 6 displays an image segmentation method based on this hybrid approach.

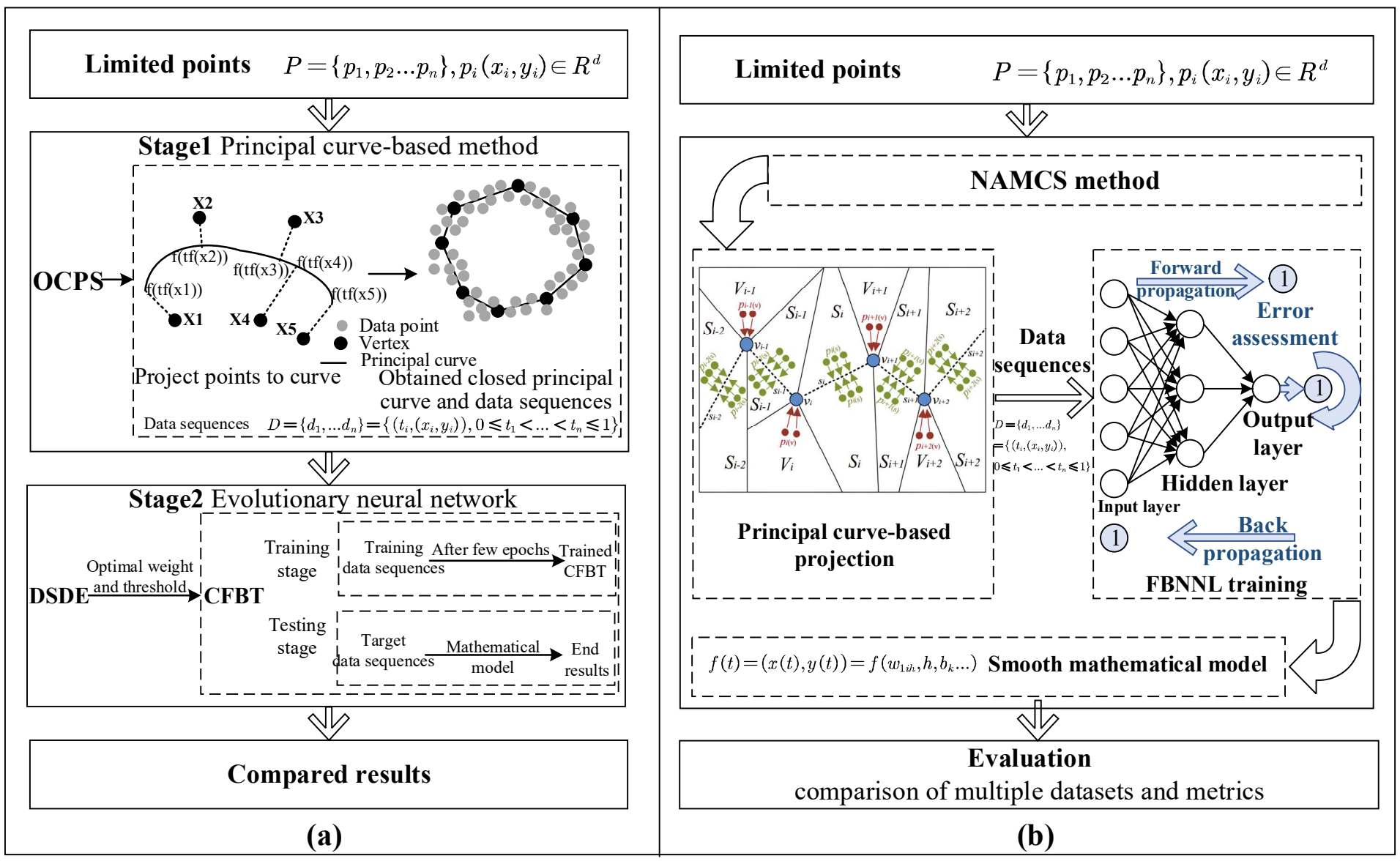

Fig. 6. Image segmentation method based on hybrid method. (a) H-ProMed [85]; (b) US Prostate Segmentation Using Adaptive Selection Principal Curve and Smooth Mathematical Model [27].

## B. *Weakly supervised learning-based medical image segmentation*

### 1) *Description of weakly supervised learning*

In image segmentation, weakly supervised learning [86] trains models using limited labeling data and weak labeling information, thus reducing dependence on manual labeling. This approach significantly cuts down the manpower and time costs associated with manual labeling. Unlike semi-supervised image segmentation methods, weakly supervised segmentation utilizes various types of weak tags, such as image-level labels, bounding box labels, scribble labels, and point labels. By using these incomplete or imprecise labels, these methods deduce pixel-level segmentation results, offering greater flexibility. They can also be combined with semi-supervised learning techniques to improve model accuracy [87].

In medical image segmentation, weakly supervised learning is effective in reducing the labor and costs associated with labeling image data. It also adeptly prevents overfitting during training, enhancing the model's robustness. Nevertheless, weakly supervised image segmentation often exhibits lower accuracy, requires extended training periods, and needs efficiency improvements [88].

*2) Applications in weakly supervised learning-based prostate segmentation*

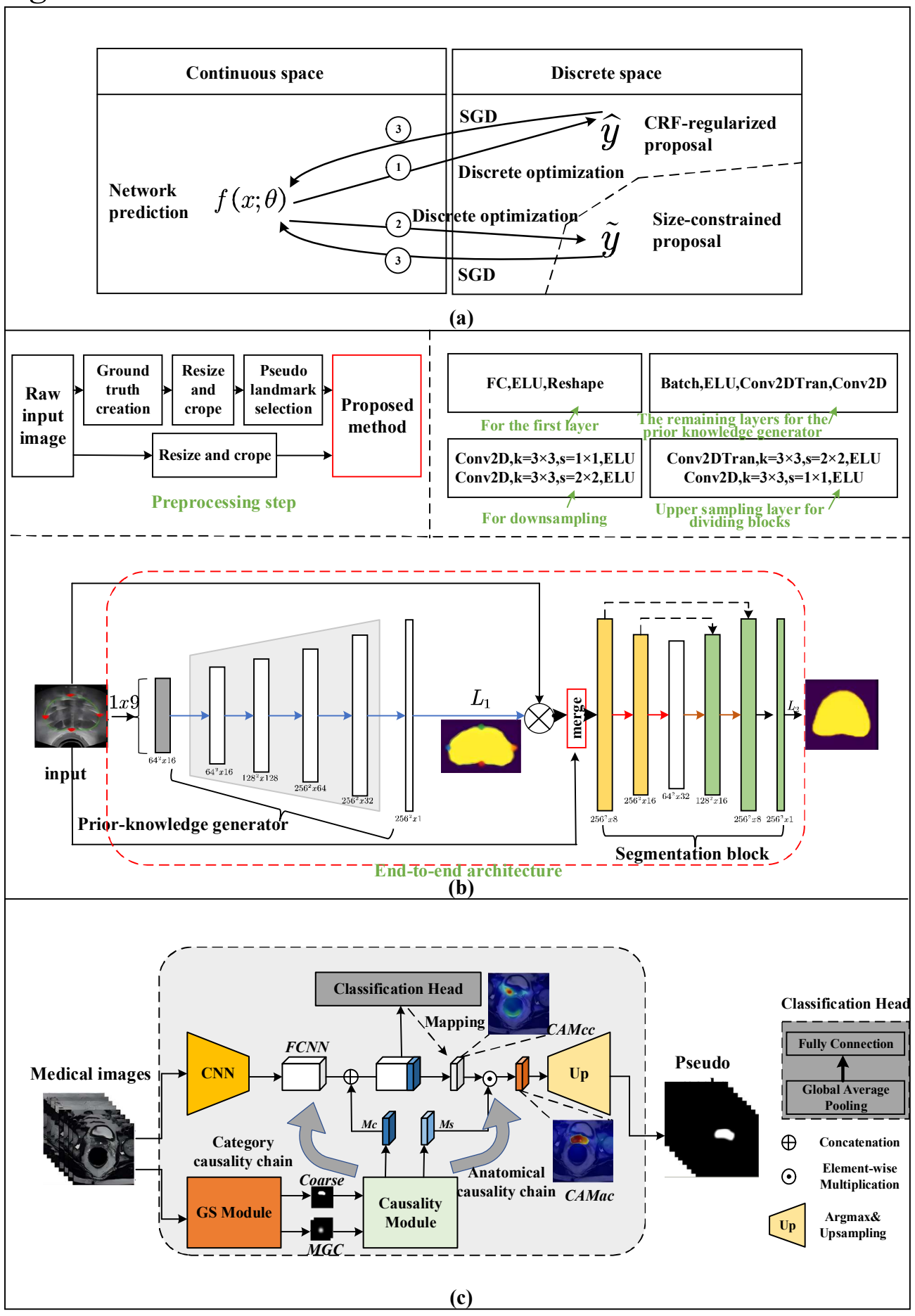

Fig. 7. Image segmentation method based on weakly supervised learning. (a) Weakly supervised discrete-continuous model diagram [89]; (b) Fast interactive medical image segmentation method based on weakly supervised deep learning [90]; (c)C-CAM architecture [91].

Peng et al. [89] developed a discrete constrained depth network for weakly annotated medical image segmentation. By integrating constraints and regularized priors, the network efficiently trains in scenarios with weak labeling. The discrete-continuous model is illustrated in Fig. 7 (a). Experiments indicate improved segmentation accuracy, constraint satisfaction, and convergence speed in prostate segmentation. Additionally, Girum et al. [90] proposed a fast interactive medical image segmentation method using a weakly supervised deep learning approach. The model diagram is shown in Fig. 7 (b). Experimental results highlight the method's efficiency in segmenting prostate clinical targets on US and CT images, achieving average Dice coefficients of 96.9 ±0.9% and 95.4 ± 0.9%, respectively, and 96.3 ±1.3% on echocardiography.

With the advancement of deep learning, semantic segmentation [91] has received considerable attention. It involves classifying each pixel, requiring training with abundant pixel-level tagging data. However, acquiring such data is labor-intensive and costly. Therefore, weakly supervised semantic segmentation (WSSS), using weak tagging, is proposed as an alternative. WSSS faces challenges like unclear target foreground and background boundaries and significant co-occurrence during training. To address these, Chen et al. [91] introduced causal activation mapping (C-CAM), depicted in Fig. 7 (c). Weakly supervised learning typically relies on large-scale centralized datasets, but federated learning (FL) provides a cross-site training approach. Zhu et al. [92] pioneered federated weak supervised segmentation (Fed WSS) and introduced the federated drift mitigation (FedDM) framework to enable segmentation model learning across multiple sites without sharing original data.

## C. Semi-supervised learning-based medical image segmentation

*1) Description of semi-supervised learning*

Semi-supervised learning [93], [94] merges the advantages and disadvantages of supervised and unsupervised learning [95]. It not only increases the accuracy of unsupervised learning but also reduces the need for labeled images, making it a popular choice in medical image segmentation. In semi-supervised image segmentation, tasks are completed by integrating various initialized splitters [96], each using different unlabeled data during training, and the results are then combined using a voting mechanism. Distinct from supervised learning, semi-supervised learning uses unlabeled data to expand the dataset, thereby reducing data collection costs. It also corrects biases in labeled data through the incorporation of unlabeled data. As a result, semi-supervised learning demonstrates improved robustness and accuracy, leading to enhanced segmentation performance [94]. However, employing semi-supervised learning involves using various unlabeled datasets to train multiple models, necessitating an integrated learning mechanism [97] for fusion. This can lead to longer training periods and reduced segmentation efficiency.

*2) Applications in semi-supervised learning-based prostate segmentation*

Recent advancements in deep learning have significantly improved the efficiency of medical data processing, including semantic segmentation [98]. These developments have provided robust solutions for automated medical image segmentation. However, training effective deep learning models usually requires a large amount of high-quality annotated data, which can be costly to collect. Nie et al. introduced ASD Net[99], a novel semi-supervised neural network based on an attention mechanism, to tackle the issue of limited data in complex networks. As illustrated in Fig. 8 (a), this method's accuracy and robustness have been validated in MRI. Additionally, Zhang et al. developed a new semi-supervised adversarial depth learning approach for semantic segmentation in 3D pelvic CT images. They used a data augmentation scheme to generate unlabeled composite data, employing an adversarial network (GANs), as depicted in Fig. 8 (b) [98]; the algorithm's details are shown in Fig. 8.

Meanwhile, Meyer et al. [100] implemented a semi-supervised learning technique called uncertainty-aware time self-learning (UATS) to bypass the costly and labor-intensive manual ground truth labeling. They integrated uncertain perceptual self-learning and time series into a new framework, enhancing the supervised deep learning model using commonly available unlabeled data. This research demonstrates the effectiveness of semi-supervised learning in contexts with limited labeled data. Li et al. [101] trained their model on prostate CT datasets using a coherent semi-supervised learning approach. They showed that semi-supervised learning's superior performance could be achieved in high-data scenarios without additional training costs, using stochastic weight averaging (SWA). Beyond CT images and MRI, TRUS is often used in prostate cancer diagnosis. However, challenges like artifacts and low resolution in US images complicate prostate US diagnostics. Xu et al. [102] proposed a shadow consistency semi-supervised learning (SCO-SSL) method incorporating two novel mechanisms: shadow enhancement (Shadow-AUG) and shadow deletion (Shadow-DROP). Dai et al. introduced an automated US segmentation method using a two-stage semi-supervised learning strategy, encompassing prostate detection and subsequent segmentation. The deep learning model's robustness was confirmed by multi-center experiments across various devices.

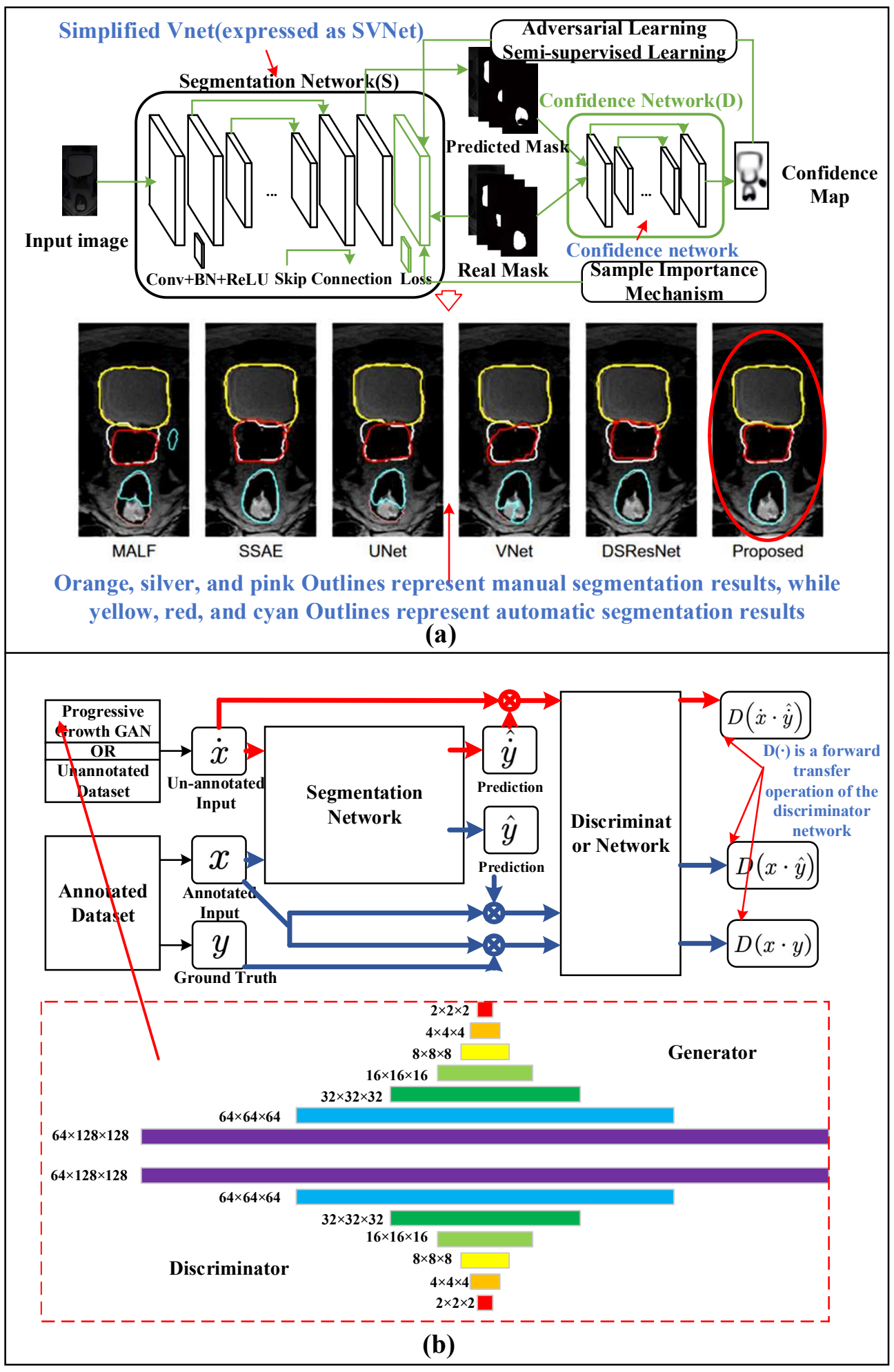

Fig. 8. Image segmentation method based on semi-supervised learning. (a) Architecture diagram of ASD Net [99]; (b) Workflow for a new semi-supervised adversarial deep learning approach [98].

### D. *Unsupervised learning-based medical image segmentation*

#### 1) *Description of unsupervised learning*

With the increasing availability of clinical data, acquiring medical images has become less challenging. However, the scarcity of experienced experts for labeling clinical data poses a significant hurdle. Unsupervised learning, in this context, plays a pivotal role in medical image segmentation. Unsupervised learning employs image structure analysis in medical image segmentation, utilizing methods like thresholding, graph cutting, edge detection, and deformation to delineate the target object's boundary in the image [103]. When image boundaries are clear, unsupervised learning segmentation is more effective, simplifying data acquisition [104].

Yet, unsupervised learning in image segmentation confronts several challenges. Firstly, the complexity of physiological structures makes obtaining accurate labels difficult, potentially leading to inaccuracies in segmentation [105]. Secondly, as it depends on image structure for segmentation, it struggles with segmenting structures having specific characteristics [106]. Lastly, segmentation outcomes from unsupervised learning often lack sufficient explanatory power, necessitating expert input for scientific interpretation of the results [107].

#### 2) *Application in unsupervised learning-based prostate segmentation*

Given that prior learning object model methods require human input for classification and are restricted to specifically tagged images, Winn and Jojic [108] introduced LOCUS (learning object class with unsupervised segmentation), which successfully learns the object class model from unlabeled images. Additionally, Liu et al. [109] developed an unsupervised segmentation method for MRI prostate using a shape prior level set. This approach determines the prostate's shape model for each subject and utilizes gradient-based techniques and morphological operators for more precise prostate segmentation. Expanding on this, Liu et al. [110] proposed an unsupervised 3D prostate segmentation method for diffusion-weighted imaging MRI, incorporating a shape prior active contour model. This method enables unsupervised segmentation in 3D MRI datasets, as shown in Fig. 9 (a) [110].

While traditional feature engineering remains an active research area, it often depends on human expertise and requires iterative adjustments. Liao et al. [111] argued for feature learning, introducing a deep learning framework illustrated in Fig. 9 (b) [111] that employs stacked independent subspace analysis (ISA) networks. These networks learn effective features hierarchically and unsupervisedly, facilitating adaptation to diverse datasets. Manual detection and delineation of the prostate in multispectral MRI data is currently time-consuming and operator-dependent, but computer-aided segmentation methods have yet to fully address these issues. To bridge this gap, Rundo et al. [112] proposed a new automatic prostate MRI segmentation method using an unsupervised learning approach based on fuzzy C-means clustering algorithm (FCM), processing multispectral T1-weighted (T1w) and T2-weighted (T2w) MRI anatomical data.

Despite the rapid development of deep learning, traditional unsupervised learning methods such as thresholding, region growth, level set, active contour, Markov random field, and clustering, continue to be used in prostate image segmentation. Conventional image segmentation [113] typically relies on heuristic rules or prior assumptions about shape, grayscale, texture, etc., to identify the prostate region. While these methods are simple, easy to implement, and require low computational resources, they suffer from drawbacks like poor robustness, sensitivity to noise and artifacts, and the need for manual parameter tuning. Fig. 9 depicts a schematic of the unsupervised learning-based image segmentation method.

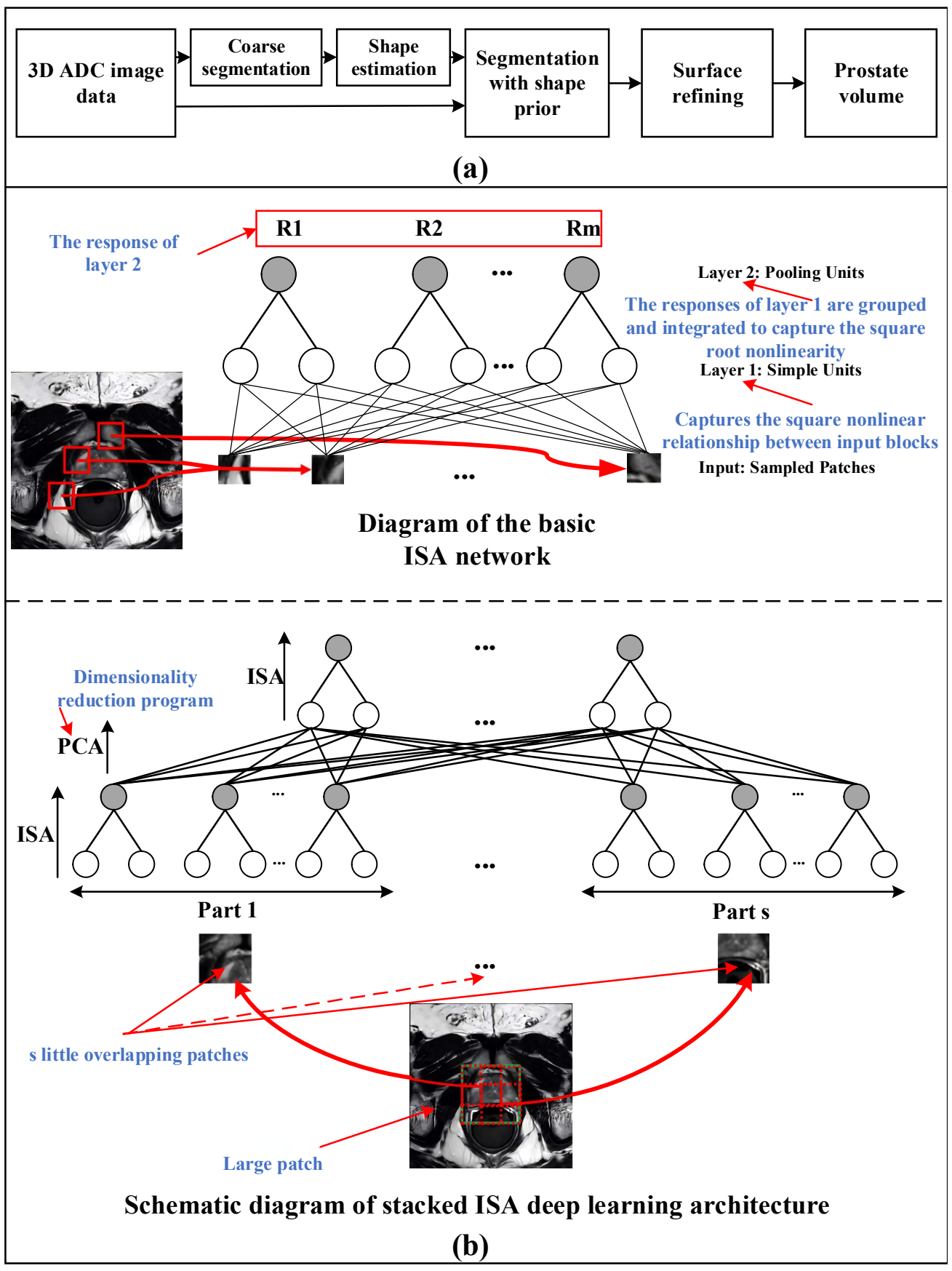

Fig. 9. Image segmentation method based on unsupervised learning. (a) Unsupervised 3D prostate segmentation method diagram based on diffusion-weighted imaging MRI [110]; (b) Schematic diagram of basic ISA network and stacked ISA deep learning framework [111].

## E. RL-based image segmentation

### 1) Description of RL in medical imaging

In image segmentation, RL [114] entails that the agent refines the segmentation quality by learning to categorize each pixel. Unlike traditional segmentation methods, RL adopts an end-to-end strategy, eliminating the necessity for manual feature design and thus improving the model's ability to generalize. RL is particularly adept at managing complex tasks and scenarios, facilitating the understanding of intricate relationships and rules within data[115]. This capability results in its superior performance in complex image segmentation tasks. RL imposes minimal constraints on labeled data, enabling effective training even with partially labeled datasets.

Nonetheless, advanced image segmentation involving the joint optimization of state, action, and reward [116], typically requires an elaborate training procedure, demanding significant computational resources and time. Additionally, the determination of appropriate state representation, action space, and reward function requires further investigation. As a result, progress in RL is relatively slow.

### 2) Applications in RL-based prostate segmentation

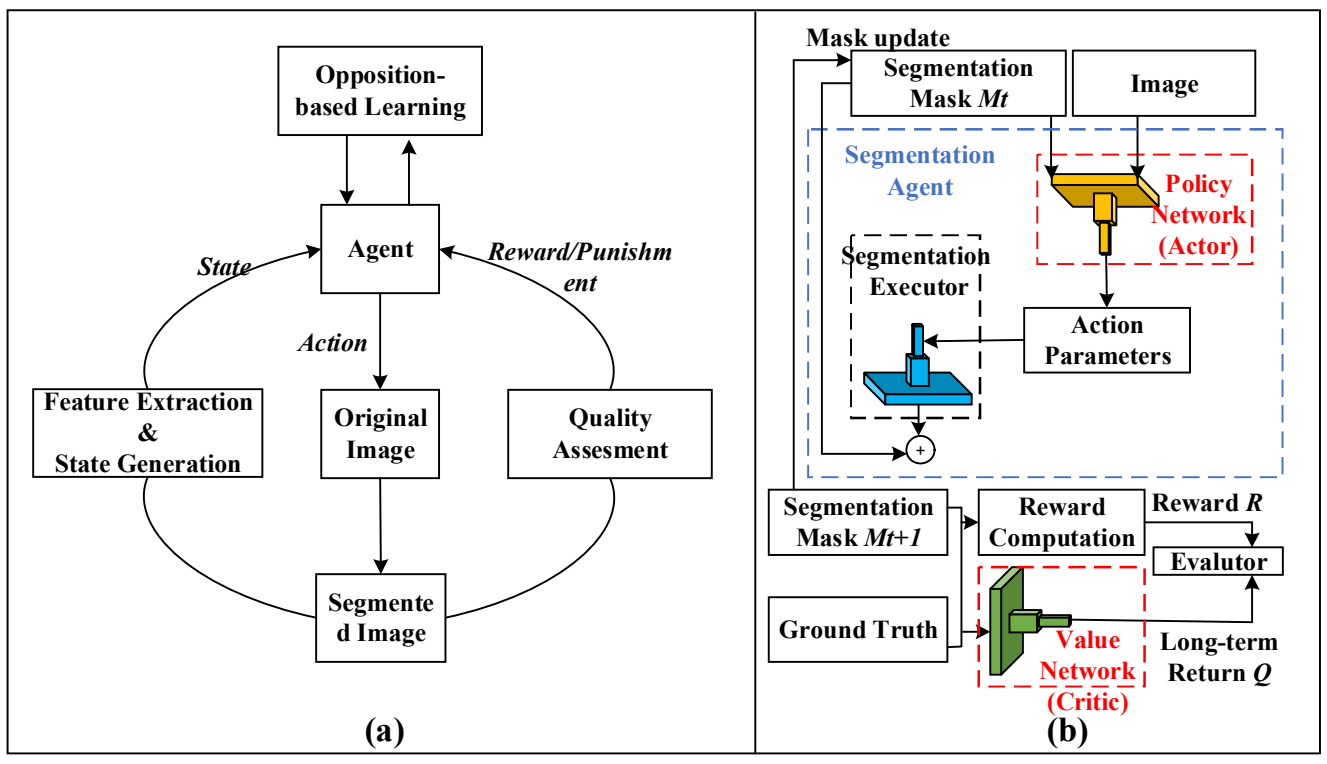

Fig. 10. Image segmentation method based on RL. (a) A general model and its components used in adversarial RL-based image segmentation methods [118]; (b) Overall architecture of automatic multi-step medical image segmentation based on deep RL [121].

Sahba et al. [117] introduced a novel RL-based method for this task. This method utilizes RL agents to adjust local thresholds and post-processing parameters. Due to its intelligent nature, RL requires only a few samples for training and can gain additional knowledge throughout the process. Subsequently, Sahba [118] developed a confrontation-based RL approach for image segmentation. As shown in Fig. 10 (a), US imaging is widely used in various medical imaging modalities. Following this, Sahba et al. [119] explored the use of RL in TRUS image segmentation. Research has shown RL's considerable promise in ultrasonic prostate segmentation. Fig. 10 illustrates a schematic of an image segmentation method utilizing RL.

In recent years, RL has been applied to a broad array of artificial intelligence challenges, including computer vision, robot control, anomaly detection, autopilot systems, computer games, and more. With the emergence of deep learning, researchers have begun integrating it with RL to tackle more complex issues. This integration has led to the development of deep RL (DRL) [120]. Tian et al. [121] use deep RL for multi-step medical image segmentation, involving training agents through deep deterministic policy gradients. This method simulates a doctor's approach to marking the region of interest (ROI) on a medical image in multiple steps, with the model's overall structure depicted in Fig. 10 (b).

## F. Verification and evaluation of models/algorithms

Prostate image segmentation is essential in diagnosing and treating prostate cancer. However, this segmentation is challenging due to factors like low contrast, high noise, and variations in prostate shape and appearance across different imaging modalities. Consequently, considerable research is focused on developing various prostate image segmentation algorithms [124]. The evaluation of these algorithms typically involves comparing their outputs with the gold standard, often the manual segmentations by experienced radiologists. However, the inherent heterogeneity of prostate anatomy and inter-observer variability during manual delineation present challenges in establishing universally accepted benchmarks. Thus, some researchers use the average of manual segmentations by different radiologists or repeated segmentations by the same radiologist as the gold standard [125].

For assessing prostate image segmentation algorithm performance, a range of standards and metrics [122], [123] are employe. These fall into two categories: qualitative and quantitative. Qualitative criteria rely on visual inspection and subjective judgment of segmentation quality, focusing on aspects like the smoothness, integrity, and consistency of prostate boundaries. In contrast, quantitative metrics are grounded in numerical analysis and objective evaluation of segmentation accuracy, employing measures such as similarity, overlap, distance, and volume error between the segmented region S and the real region G.Some commonly employed quantitative metrics for prostate image segmentation are the Dice similarity coefficient (DSC), Jaccard index (JI), Haus Dorff distance (HD), among others. These metrics offer objective and thorough evaluations of prostate image segmentation algorithms. However, they have limitations and challenges. For instance, lack scale or rotation invariance, and may not adequately reflect the clinical significance of segmentation errors. Therefore, when assessing prostate image segmentation algorithms, it is imperative to use a variety of metrics. Furthermore, the use of standardized datasets and protocols for equitable comparison and benchmarking of different algorithms is vital [126]. For effective performance assessment of segmentation methods, they should be compared using common datasets. However, the development of various methods tailored to specific application scenarios makes it challenging to employ standardized metrics for quantitative comparison across the same dataset.

## IV. Discussion and Summary

### A. Discussion

While automatic prostate segmentation methods reduce human error and improve effectiveness, many of these algorithms are complex, computationally demanding, and require faster segmentation speeds. Deep learning automatically extracts complex features from large datasets, providing more precise image region representations [127], thus enhancing image segmentation performance and efficiency. Compared to traditional methods, deep learning necessitates longer training, larger datasets, and greater processing power, benefiting from expanded data capacities and advancements in computer hardware. Recently, deep learning has achieved significant progress in computer vision, leading to a proliferation of deep learning-based image segmentation methods, like U-Net [47] and Transformer [128].

Considering the nature of medical imaging technology (Part 2), single-modality imaging has its limitations, particularly in capturing rich and comprehensive image information. Additionally, image quality is often affected by environmental factors like lighting and noise. Moreover, different pathological conditions may appear similar in the same modality, complicating differentiation. Hence, single-modality images often fail to ensure accurate and robust segmentation. In contrast, multi-modality approaches provide more comprehensive image data. By integrating information from various modalities, they enhance the accuracy of prostate tissue and lesion segmentation, better addressing complexity and uncertainty in medical images, and improving segmentation model performance and robustness. However, multi-modality segmentation methods face challenges, such as the need for increased computational resources, leading to higher segmentation costs. Additionally, inconsistencies in data and labels across modalities complicate effective information fusion, requiring well-designed fusion strategies for optimal use of each modality's data. Despite these challenges, multi-modality image segmentation remains a promising research area.

Literature research reveals that although many three-dimensional ultrasonic prostate segmentation methods have been proposed, a substantial number of these methods still rely on two-dimensional techniques. The primary reason is that in most hospitals, US prostate images are obtained by inserting a two-dimensional probe into the rectum through the anus, with doctors manually adjusting the probe to capture standard cross-sectional and longitudinal views of the prostate. Factors like prostate size and shape, slimitations due to the anal canal's length and the tightness of the anal sphincter, hinder setting consistent section depths manually. Consequently, the transition to using transrectal three-dimensional probes for acquiring three-dimensional US prostate data, and thereby developing three-dimensional segmentation methods, is anticipated to be a trend in the coming years.

The evolution of medical image segmentation technology has progressed from manual, to semi-automatic, and then to fully automatic segmentation. While manual segmentation is simple, straightforward, and highly accurate, it is time-consuming and subject to subjective biases, demanding extensive expertise from physicians and entailing laborious tasks. To balance efficiency and accuracy, semi-automatic segmentation methods emerged. These methods significantly improve speed compared to manual segmentation and maintain high accuracy, yet they still require some manual involvement. Automatic segmentation, developed to eliminate manual intervention, forms the basis for automatic and precise measurement of regions of interest in images.

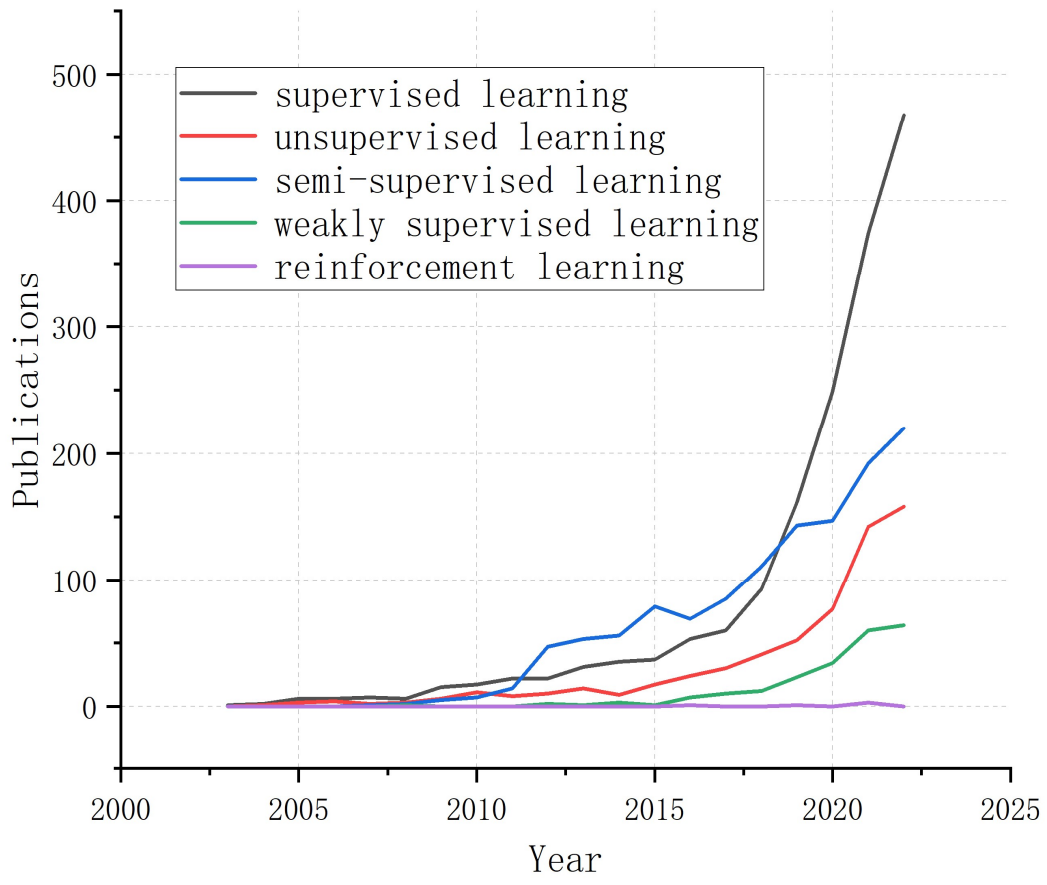

Fig. 11. Number of literature in the PubMed database from 2003 to 2022 with the keywords "image segmentation" and "supervised/unsupervised/semi-supervised/weakly supervised/RL" in the abstract or title.

Moreover, according to literature research (Fig. 11), a significant portion of image segmentation tasks currently relies on supervised learning, despite the high cost of labeling quality datasets. Due to limited labeled data, research in unsupervised, semi-supervised, and weakly supervised learning within image segmentation has gained momentum in recent years. Image segmentation using RL is challenging, time-intensive, and constrained by technical issues, resulting in limited research in this area. However, this method holds potential for application in more complex and diverse scenarios. As RL continues to mature, its applications are expected to become more widespread.

## *B. Summary*

This paper offers a detailed analysis of recent artificial intelligence-based automatic prostate segmentation methods. It explores their inherent features and importance in medical imaging and image segmentation, addresses the verification and performance evaluation processes of these methods, and identifies the current challenges in this field. The paper concludes by discussing future research directions and trends in AI-based automatic prostate segmentation.

We draw the following conclusions:

- Advancements in computer hardware and the growth in dataset capacity have facilitated the development of various deep learning methods, demonstrating enhanced segmentation performance. Technologies such as deep learning and machine learning have emerged as focal points and challenges in recent research.
- Traditional image segmentation algorithms can enhance the efficiency of deep learning models in medical image segmentation [129].
- There is a trend towards expanding from two-dimensional imaging to three-dimensional or even four-dimensional modalities, and a shift from semi-automatic to fully automatic segmentation.
- Given the limitations of single-modality images, the transition to multi-modality image segmentation appears promising.
- In light of restricted annotated data availability, there is growing interest in exploring unsupervised learning and applying RL. In conclusion, the precision detection and treatment of prostate cancer are advancing positively.